\pdfoutput=1

\documentclass{elsart}%
\usepackage{amsmath}%
\usepackage{amsfonts}%
\usepackage{amssymb}%
\usepackage{graphicx}
\begin{document}
%
\begin{frontmatter}%

\title
{ Equilibrium (Zipf) and Dynamic (Grasseberg-Procaccia) method  based analyses of human texts.
  A comparison of natural (english) and artificial (esperanto) languages.   \\ }%

\author{ M. Ausloos}%

\address{
 GRAPES, U. Liege, B5 Sart-Tilman, B-4000 Liege, Belgium}%

\begin{abstract}
A comparison of  two english texts from Lewis Carroll, one (Alice in wonderland), also translated into esperanto, the other (Through a looking glass) are discussed in order to observe whether natural and artificial languages  significantly differ from each other. One dimensional time series like signals are constructed using only word frequencies (FTS) or word lengths (LTS). The data is studied through (i) a Zipf method for sorting out correlations in the FTS  and (ii) a Grassberger-Procaccia (GP) technique based method for finding correlations in LTS.     Features  are compared :  different power laws  are observed with characteristic exponents for the ranking properties, and the {\it phase space attractor dimensionality}.  The Zipf exponent can take values much less than unity ($ca.$ 0.50 or 0.30) depending on how a sentence is defined.  This non-universality is conjectured to be a measure of the author $style$. Moreover the attractor dimension $r$ is a simple function of the so called phase space dimension $n$, i.e., $r = n^{\lambda}$, with $\lambda = 0.79$. Such an exponent should also conjecture to be a measure of the author $creativity$. However, even though there are quantitative differences between the original english text and its esperanto translation, the qualitative differences are very minutes, indicating in this case a translation relatively well respecting, along our analysis lines, the content of the author writing.

\end{abstract}


\end{frontmatter}%

\section{Introduction}

Human languages are systems usually composed of a large number of internal components (the words, punctuation signs, and blanks in printed texts) and rules (grammar). Relevant questions pertain to the life time, concentration, distribution, .. complexity of these and their relations between each others.  
Thus human language is a new
emerging field for the application of methods from the physical sciences in order to achieve a deeper
understanding of linguistic complexity \cite{Schulzereview06,staufferreview,Schulzereview07,schulzestauffer2,schulzestauffer3}.
Language distributions, competitions, life durations, ... have become an active field  of research in statistical  physics indeed since  \cite{schulzestauffer2,schulzestauffer3,viviane,1}, where usual techniques  based on non-equilibrium considerations \cite{fujita}, and agent based models are already much applied

One should distinguish two main frameworks. On one hand, language developments seem to be understandable through competitions, like in Ising models, and in self-organized systems. Their diffusion seems similar to percolation and  nucleation-growth problems  taking into account the existence of different time scales, for  inter- and intra- effects. The other frame is somewhat older and originates from  more classical linguistics studies; it pertains to the content and meanings \cite{klinkenberg,chomsky}. This latter case is of interest here and the main subject of the report, within a statistical physics framework.

 Concerning the internal structure of a text, supposedly characterized by the language in which it is written,  it is well known that a text can be mapped into a signal, of course first  through the alphabet characters. However it can be  also reduced to less abundant symbols through some threshold, like a time series, which can be a list of +1 and -1, or sometimes 0. Thereafter one could apply at this stage many techniques of signal analysis.
 
In fact,  laws of text content and structures have been searched for  a long time by e.g. Zipf and others \cite{Zipf1,Zipf2,yule,meadowartificial,powers} through the least effort (so called ranking) method. The technique is now currently applied in statistical physics as a first step to obtain, when they exist, the primary scaling law. It has been  somewhat a surprise that the number of words $w(h)$ which  occurs $h$ times in a text is such that   $w(h)\sim 1/h^\gamma$, where $\gamma \sim 2$, while  the rank $R$ of the words according to their frequency $f$ behaves like another power law  $ f \sim R^{-\zeta}$ where the exponent $\zeta$ is quasi always close to 1.0 \cite{west}. Some thought has been presented to explain so, based on constrained correlations \cite{modelzipf1,modelzipf2}.  Another distribution has been studied, i.e. the distribution of word lengths in a text.  Whence two features can be looked for  (i) word frequencies (FTS) or  (ii) word lengths (LTS). We hereby consider that in physics terms they represent different measures of the system : the first one leads to characterizing  the spanned phase space through a measure, - it is a static-like, equilibrium approach, obtained  $after$ the text is finalized, while the second rather contains a time $evolution$ aspect : it takes more time to pronounce (or read) a long word than a small one. Whence another technique of analysis than the Zipf one should be put forward. We implement the  Grassberger-Procaccia (GP) technique   for finding ''time correlations'' in the text through the analysis of LTS, as a signal spanning some attractor in a space on an  {\it a priori} unknown dimension.
 
 Obviously there are many ways to map a text onto a time series, but in the present study  the above two series are only considered, due to their physical meaning  which can be thought to be implied in the mapping. 

 No need to recall the many communications in which a comparison of the properties of such ''time signals'' has   been presented, - sometimes even (very) artificial so called languages \cite{meadowartificial} have been discussed, like those used for simulation codes on computers \cite{fortran}. Comparison of different truly human languages arising from apparently different origins or containing  different signs has also been made, e.g. beside english, one can find references about greek \cite{LTSpanos,greek1,greek2}, turkish \cite{turkish}, chinese \cite{chinese}, ...   
``Linguistic time series''  have often studied at a letter or word level \cite{39,41,42,42b} or as in Montemurro and Pury \cite{42,42b} at a ÔÔfrequency
mappingÕÕ,  similar though not identical to the one described below. Others have considered Zipf law(s) at the sentence level \cite{sentence,ebelingsentence}, - a few sometimes strangely neglecting the punctuation \cite{irish,kwapien}.

  Esperanto is an artificially and somewhat recently constructed language \cite{esperanto}, which was intended to be an easy-to-learn lingua franca. Previous statistical analyses seem to indicate that esperanto's statistical proportions are similar to those of other languages \cite{manaris}.  It was found that esperanto's statistical proportions resemble mostly those of German and Spanish, and somewhat surprisingly least those of French and Italian. By the way, english seems to be an intermediary case \cite{powers}. Yet there are quantitative differences : English  contains  ca.   1 M  words \cite{monitor}, esperanto 150 k words \cite{plena}.  
   Other artificial languages exist,  like that of the Magma \cite{magma} and Urban Trad  \cite{urbantrad} music groups, the latter specifically designed for song competition, i.e. the eurovision contest \cite{eurovision}. Like in e.g.  rap music lyrics or french $verlan$, the thesaurus is rather of limited size in all these cases.

To  my  knowledge few comparisons exist on texts translated from one to another language \cite{fnote1,memoireUCL,newIJMPC,kanter}, in particular into artificial  languages.
 We present below an original consideration in this respect,  the analysis and results about a translation between one of the most commonly used language, i.e. english, and a relatively recent language, i.e. esperanto. 
 
 The text to be used was chosen for its wide diffusion, freely available from the web \cite{Gutenberg} and as a representative one of a famous scientist, Lewis Caroll, i.e. {\it Alice in wonderland} (AWL) \cite{carollAWL}. Moreover knowing the special (mathematical) quality of this author's mind, and some, as I thought {\it a priori},  possibly special way of writing, a bench mark has been chosen for comparison, i.e.  {\it Through a looking glass} (TLG)  \cite{carollTLG}; - alas to my knowledge only available in english on the web \cite{footn2}. Yet this will allow us to discuss whether the difference, if any, between esperanto and english, are apparently due to the translation or on the contrary to the specificity of this author's work. It might be also expected that one could  observe whether some style or vocabulary change has been made between two texts having appeared at different times : 1865 and 1871, or not.  Previous work on  the english AWL version should be mentioned \cite{powers}, where the discussion mainly pertains on corpus size effect on the validity of Zipf law, but where is emphasized a relevant ingredient to be taken into account in discussing   most written texts, i.e. a mixing of oral and descriptive accounts.
  
 In Sect. 2, a few elementary facts and basic statistics on these texts are  presented;  the methodology is   briefly exposed, i.e. as one  recalls (i) two simple ways to map texts into $ signals$, i.e.,   the frequency
time series (FTS) and the  (word) length time series (LTS) , (ii)  the Zipf ranking technique,  (iii) the Grassberger-Procaccia (GP) method \cite{GPprl,GPphd}  used for finding correlations. 
Similar techniques for comparing english and greek texts, but not from a translation point of view can be found in \cite{LTSpanos}; however the published work contains a few annoying (misprints or) defects which induces us to reformulate the techniques when applied to the present problem. In Sect. 3,  the results are presented : (i) a Zipf analysis on  the frequency
time series (FTS), (ii)  a GP analysis for the  (word) length time series (LTS).  The results  are discussed in Sect. 4.



 \section{Data and Methdology}
 
 For these considerations  two texts here above mentioned   and one translation have been selected and downloaded from a freely available site \cite{Gutenberg}, resulting obviously into three files. The chapter heads  have been removed. All   analyses are carried out over this reduced file  for each text.  Basic statistics, like the number of words, the longest sentence, ... are given in Table \ref{tab-1} for each text, and chapters. A few facts attract some attention
 \begin{enumerate}
 \item the number of dots is {\it much smaller} in AWL$_{\mbox{eng}}$ than in AWL$_{\mbox{esp}}$ and also in   TLG$_{\mbox{eng}}$ 
 \item automatically the longest sentence occurs in AWL$_{\mbox{eng}}$ with many more characters
 \item the longest sentence in AWL$_{\mbox{esp}}$  occurs between commas
 \item the number of semi columns is very small in  TLG$_{\mbox{eng}}$
\item  the longest sentence ever occurs in TLG$_{\mbox{eng}}$  between semi-columns
\item there are {\it very few } exclamation marks in AWL$_{\mbox{esp}}$ 
\item but a long sentence is then found between these in such a work
\item more importantly the number of sentences is much smaller in AWL$_{\mbox{eng}}$ than in AWL$_{\mbox{esp}}$. 
\end{enumerate}
 

\begin{table}
\tiny{
\begin{center}
\begin{tabular}{|c|c|c|c|} \hline
&	AWL$_{\mbox{eng}}$	&	AWL$_{\mbox{esp}}$	&	TLG$_{\mbox{eng}}$	\\ \hline \hline
Number of   words	&	27342	&	25592	&	30601	\\ \hline 
Number of  different words	&	2958 	&	5368 	&	3205\\ \hline
Number of characters	&	144927	&	154445	&	164147	\\ \hline \hline
Number of punctuation marks	&	4481 	&	4752 	&	4828 \\ \hline
Number of ''sentences'' &	1633 	&	2016 	&	2059 \\ \hline \hline
Words in chap. 1& 2194  & 1858 & \\ \hline 
Different words in chap. 1&652 & 853 & \\ \hline \hline
Words in chap. 2& 2188 & 1915 & \\ \hline 
Different words in chap. 2& 665 &829&\\ \hline \hline
Number of dots	&	979	&	1545	&	1315	\\ \hline
Longest ''sentence''	&	1669	&	825	&	864	\\ \hline \hline
Number of commas	&	2419	&	2324	&	2441	\\ \hline
Longest ''sentence''	&	373	&	1170	&	368	\\ \hline \hline
Number of semi columns	&	195	&	207	&	72	\\ \hline
Longest sentence	&	6624	&	6043	&	12501	\\ \hline \hline
Number of columns	&	234	&	205	&	256	\\ \hline
Longest sentence	&	4586	&	5576	&	3145	\\ \hline \hline
Number of question marks	&	203	&	205	&	254	\\ \hline
Longest sentence	&	6323	&	5581	&	5212	\\ \hline \hline
Number of exclamation marks	&	451	&	266	&	490	\\ \hline
Longest sentence	&	4388	&	6249	&	4016	 \\ \hline \hline
\end{tabular}
\end{center}
\caption{Basic statistical data for the three texts of interest; in each case the longest sentence is measured in terms of the number of characters (not in terms of words)}
}
\label{tab-1}
\end{table}

 \begin{figure}
 \begin{center}
\hspace{-0.2cm}
\includegraphics[width=2.8in]{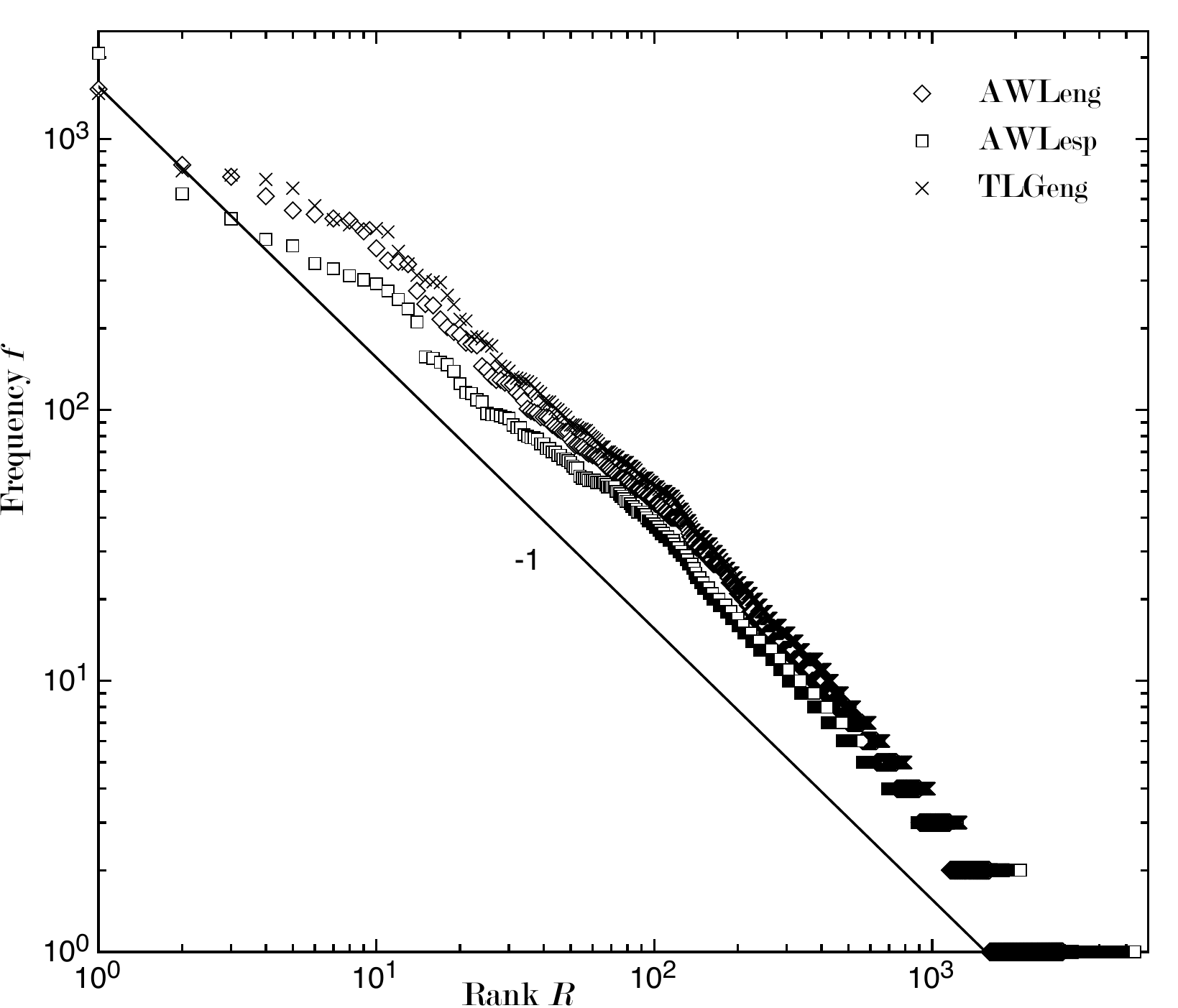}
\vspace{-0.2cm}
\caption{\label{1} Zipf (log-log) plot of the frequency of words in the three texts of interest AWL$_{\mbox{eng}}$, AWL$_{\mbox{esp}}$ and TLG$_{\mbox{eng}}$. The usual ($\zeta=1$) exponent is indicated. A Zipf-Mandelbrot law fit for $2 \le R \le 1000$ is not shown but is discussed in the main text; see also table III }
\end{center}
\end{figure}

Let us now search for correlations in the texts through both ways  of constructing a time series from such documents of  e.g. $M$ words:
\begin{enumerate}
 \item   
 Count the frequency $f$ of appearance of each word in the document. Rewrite the text such that at each  ''appearance'' of a word, the word is replaced by its   frequency such that one obtains a time series $f(t)$.   Such a time series is called a ''frequency time series'' (FTS).


\item  Count the  number  $l$  of letters of each word located in the text successively at  the $time$ 
$t $= 1, for the first word,   at time $t$ = 2, for the second, etc. Construct a time series $l(t)$. Henceforth,   such a time series is called a ÔÔlength time seriesÕÕ (LTS).
 \end{enumerate}

 
 When applied e.g., to economic (financial) signals  \cite{maki2,PhBMA2,PhBMA418,surya},  each frequency $f$ and word  length $l$ are analogous to  the price of a share or the volume of a transaction.  A (scaling or) power law is  then often observed, i.e. when correlated sequences exist, leading to $\zeta_{m,k}$ values quite different from 1. 

\begin{figure}
 \begin{center}
\hspace{-0.2cm}
\includegraphics[width=2.7in] {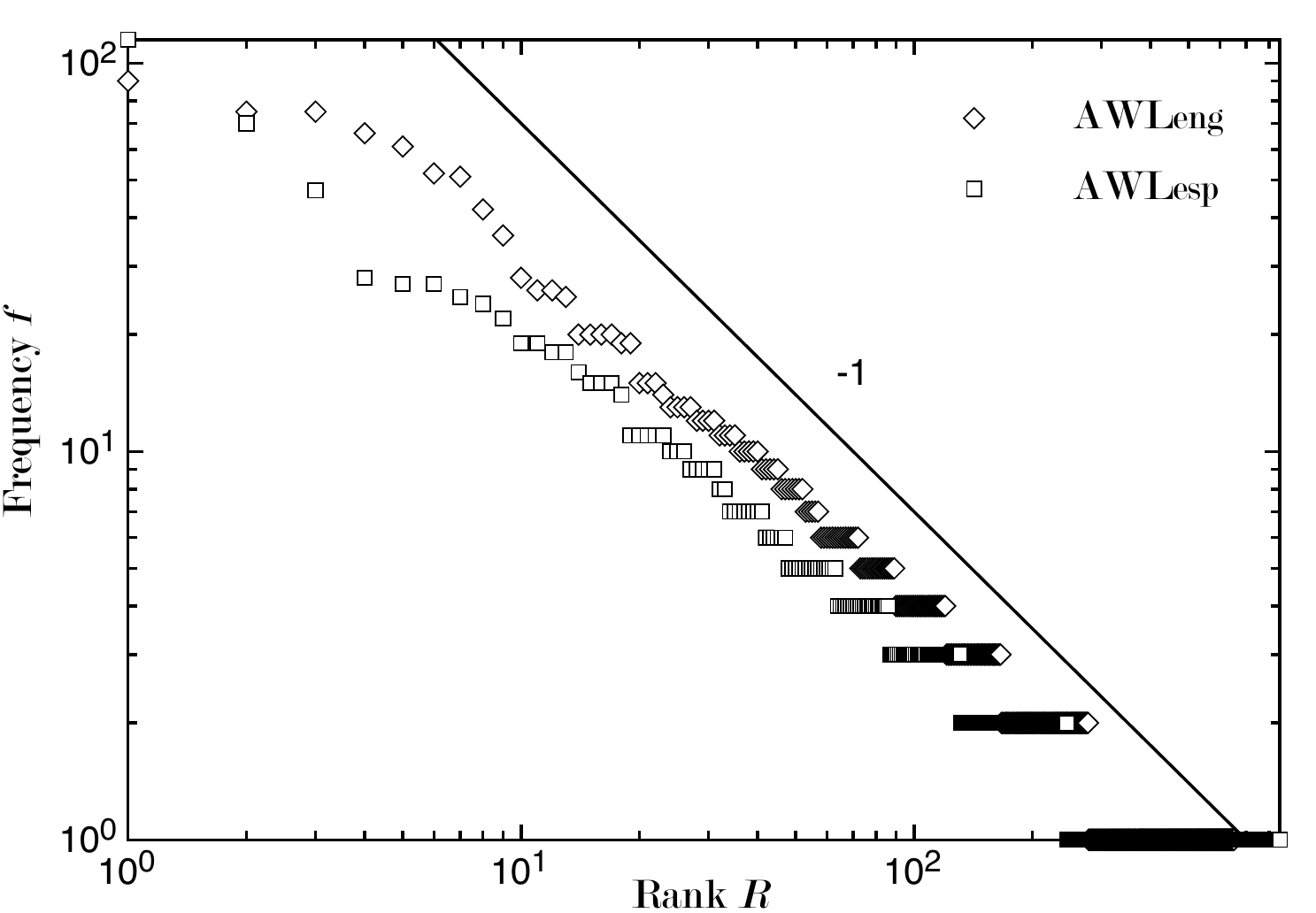}
\includegraphics[width=2.7in] {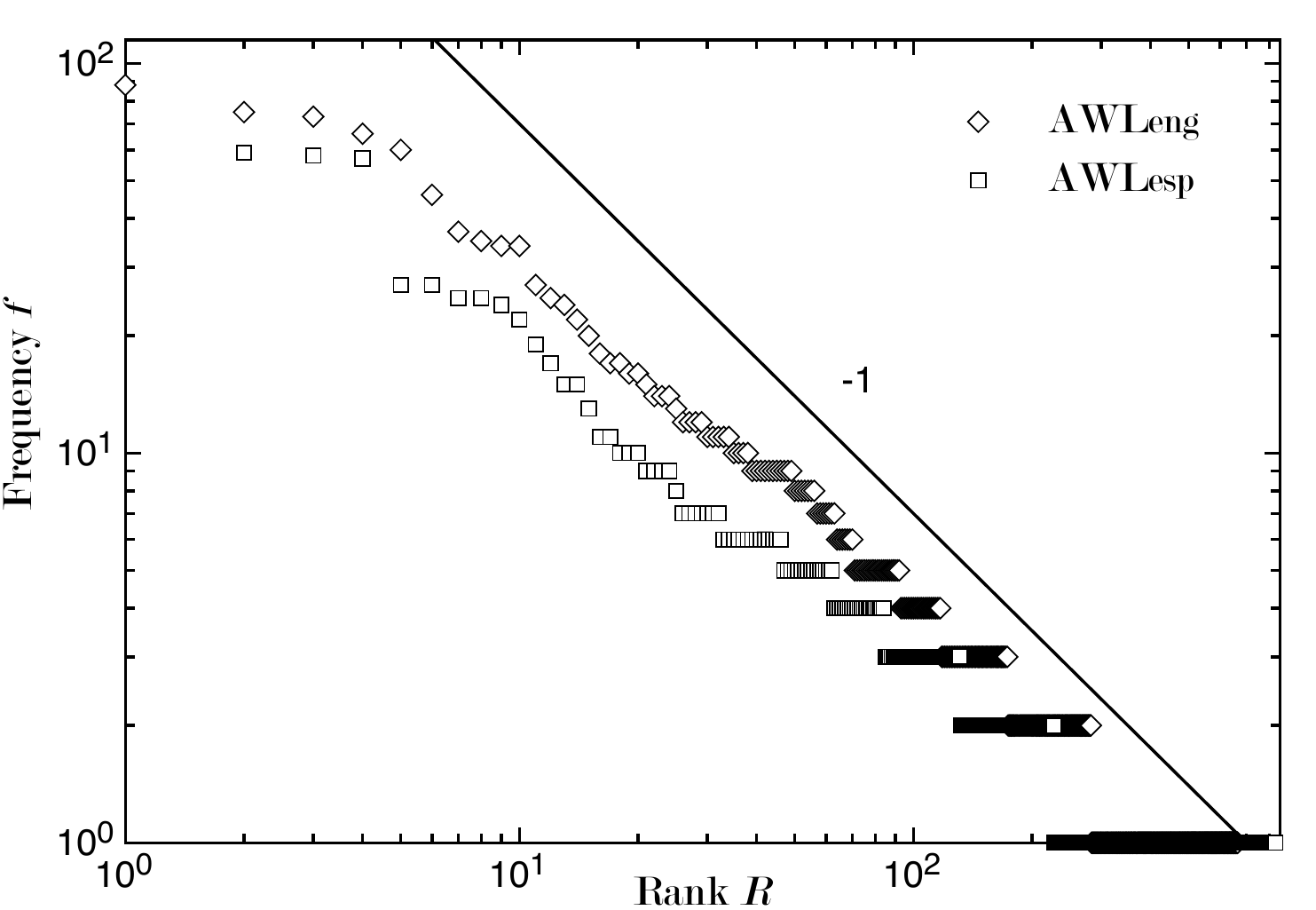}
\vspace{0.2cm}
\caption{\label{2}  Zipf (log-log) plot of the frequency of words in  (a) chapter 1, (b)   chapter 2,  for  the texts of interest, i.e. AWL$_{\mbox{eng}}$ and AWL$_{\mbox{esp}}$. The ''usual''  1 exponent is indicated. A Zipf-Mandelbrot law fit for $ R \le 200$ is not shown but is discussed in the main text; see also table III  }
 \end{center}
\end{figure}

These two sorts of time series are thereby analyzed along one of the  two mentioned techniques, one being more pertinent than the other as outlined here above. Let us discuss them briefly.

 
 
   \subsection{Zipf method}
 
A large set of references on Zipf's law(s) in natural languages can be found in \cite{httpZipf}. The idea has been applied to many various complex signals or ''texts'', -   signals, translated through a number $k$ of characters characterizing an alphabet, like, among many others,  for time intervals between earthquakes \cite{suzuki}, DNA sequences \cite{dna
} or for financial data \cite{maki2,PhBMA2,PhBMA418} along the lines of econophysics.


\begin{figure}
 \begin{center}
\hspace{-0.2cm}
\includegraphics[width=2.7in] {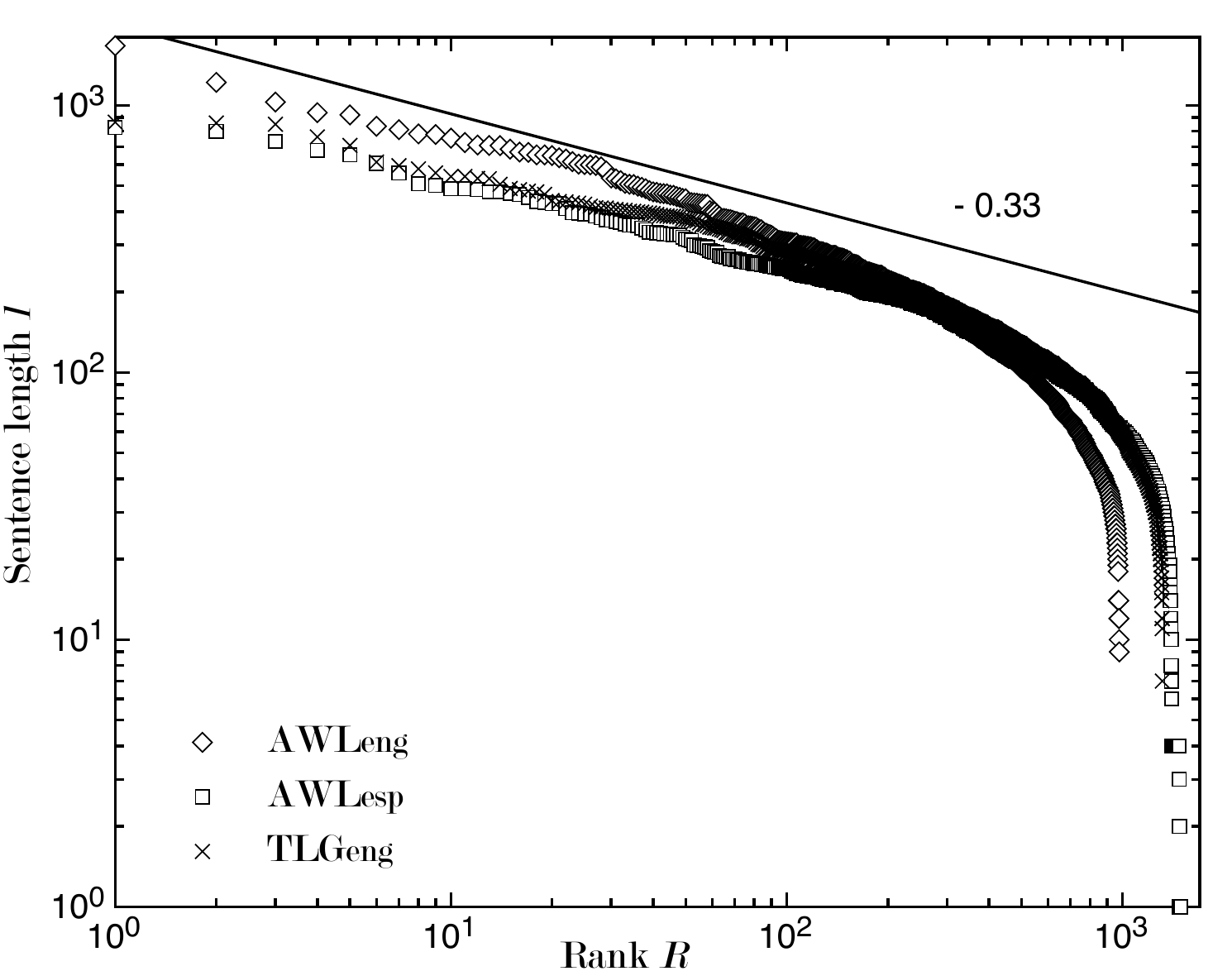}
\vspace{-0.1cm}
\includegraphics[width=2.7in] {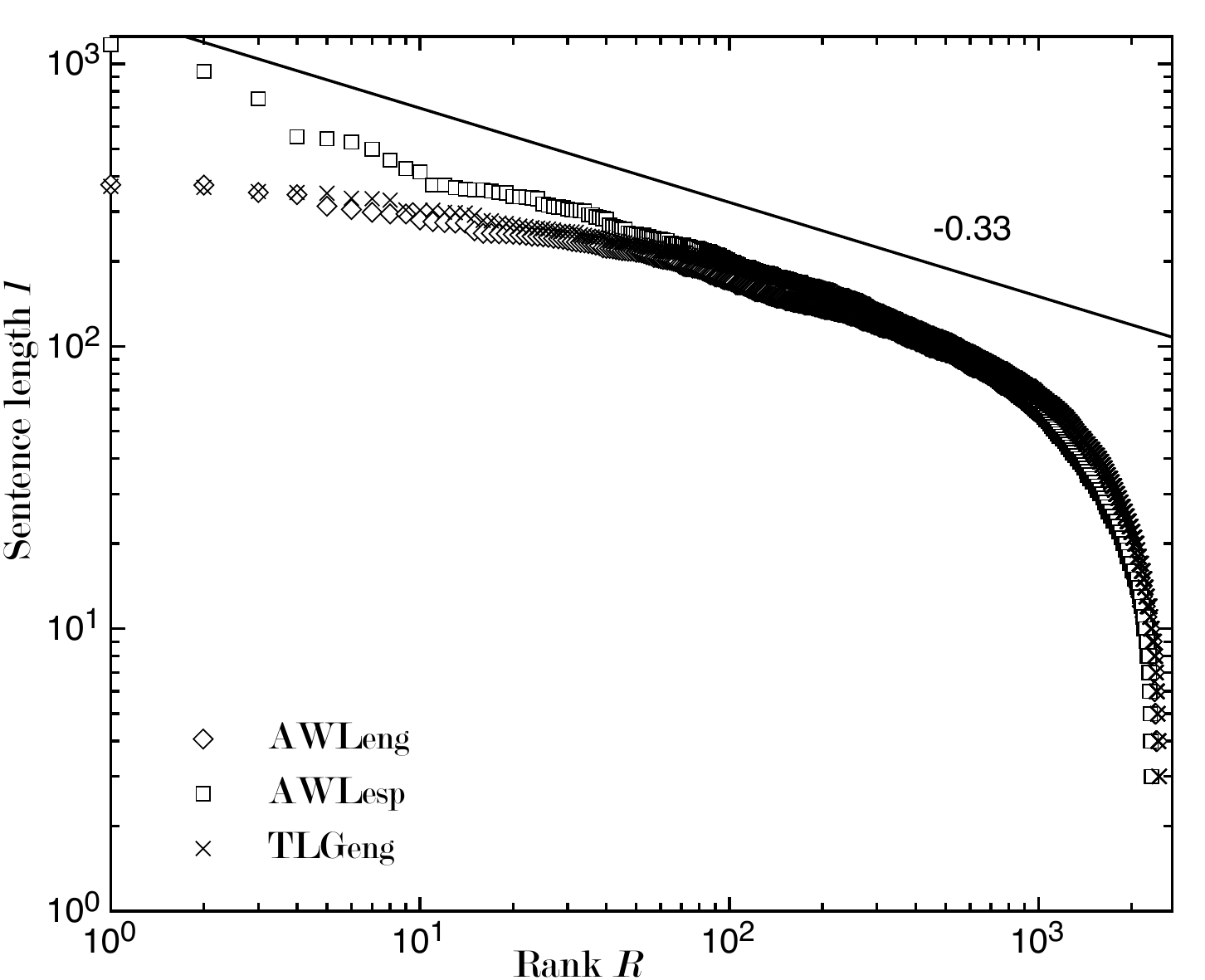}
\vspace{-0.01cm}
\caption{\label{3} Zipf (log-log) plot of the FTS of sentence lengths, as separated by (a) dots, (b) commas,  in the three texts of interest AWL$_{\mbox{eng}}$, AWL$_{\mbox{esp}}$ and TLG$_{\mbox{eng}}$. The  0.33 exponent of the corresponding Zipf law is indicated as a guide to the eye. A Zipf-Mandelbrot law fit for $ R \le 200$ is not shown but is discussed in the main text; see also Table III}
\end{center}
\end{figure}

\begin{figure}
 \begin{center}
\hspace{-0.2cm}
\includegraphics[width=2.7in] {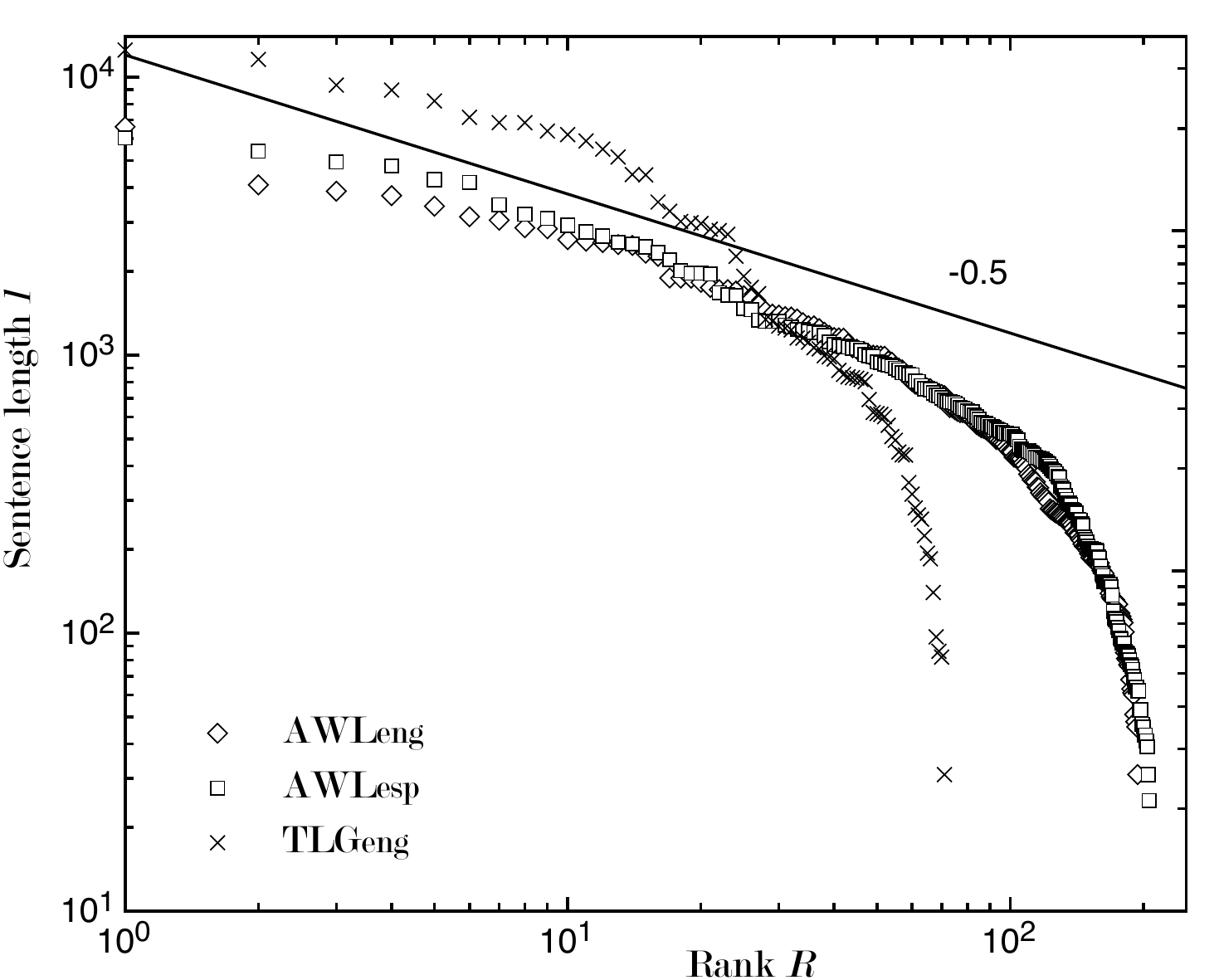}
\vspace{-0.1cm}
\includegraphics[width=2.7in] {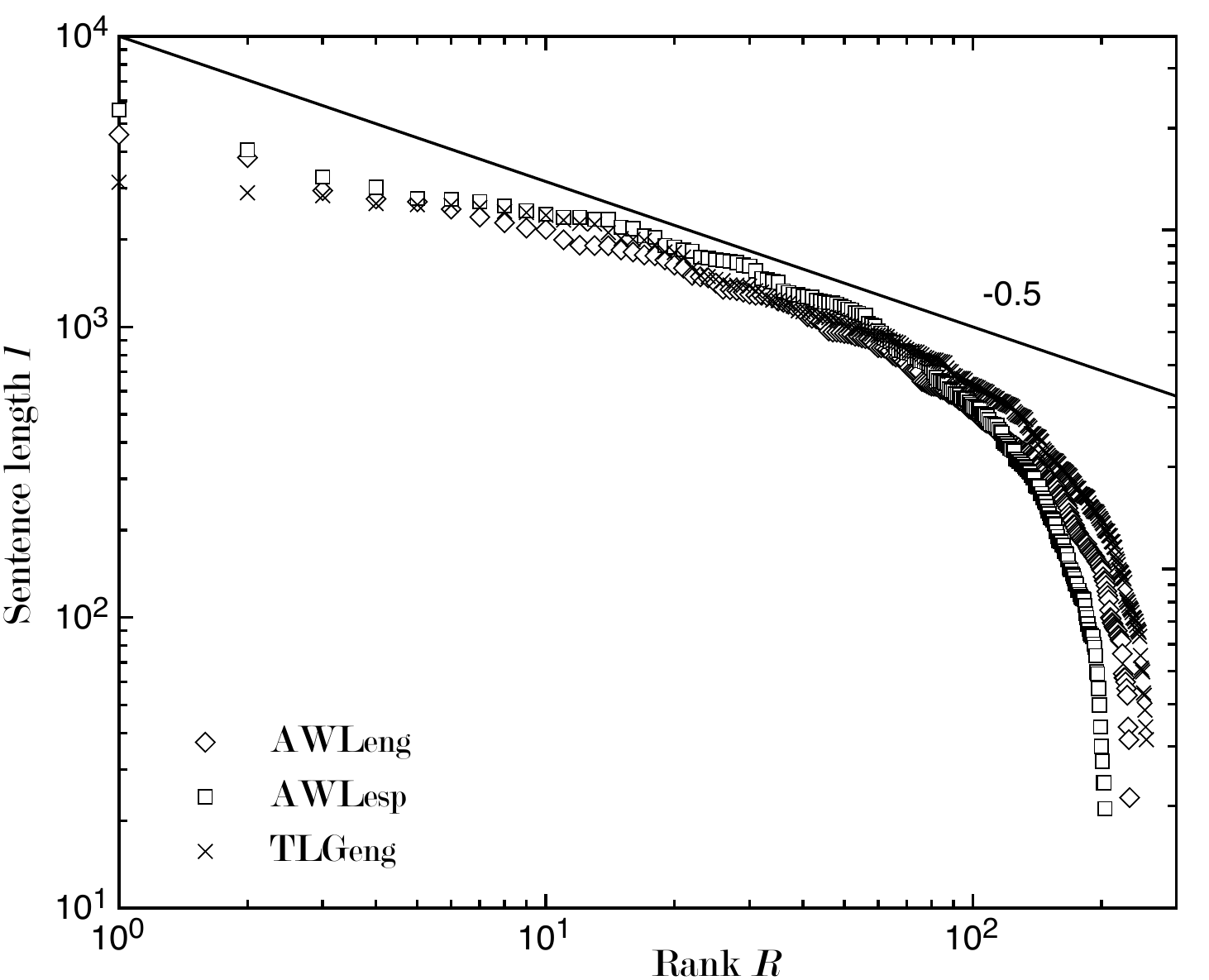}
\vspace{-0.01cm}

\caption{\label{4} Zipf (log-log) plot of the FTS of sentence lengths, as separated by (a) semicolums, (b) columns,  in the three texts of interest AWL$_{\mbox{eng}}$, AWL$_{\mbox{esp}}$ and TLG$_{\mbox{eng}}$ . The  0.50 exponent of the corresponding Zipf law is indicated as a guide to the eye. A Zipf-Mandelbrot law fit for  various $R$ ranges  is not shown but is discussed in the main text; see also Table III}
\end{center}
\end{figure}

 \begin{figure}
\hspace{-0.1cm}
\includegraphics[width=2.5in] {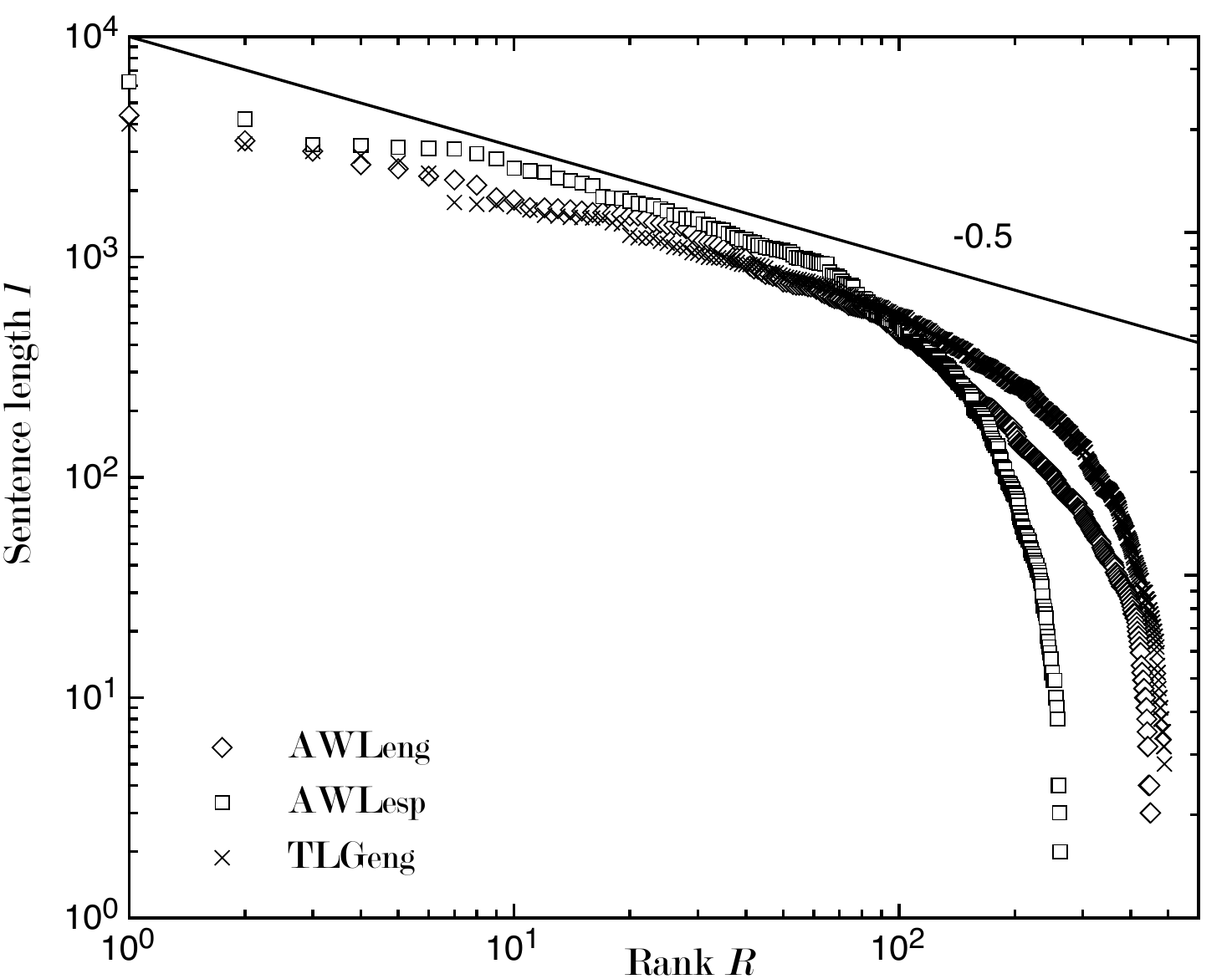}
\vspace{-0.01cm}
\includegraphics[width=2.5in] {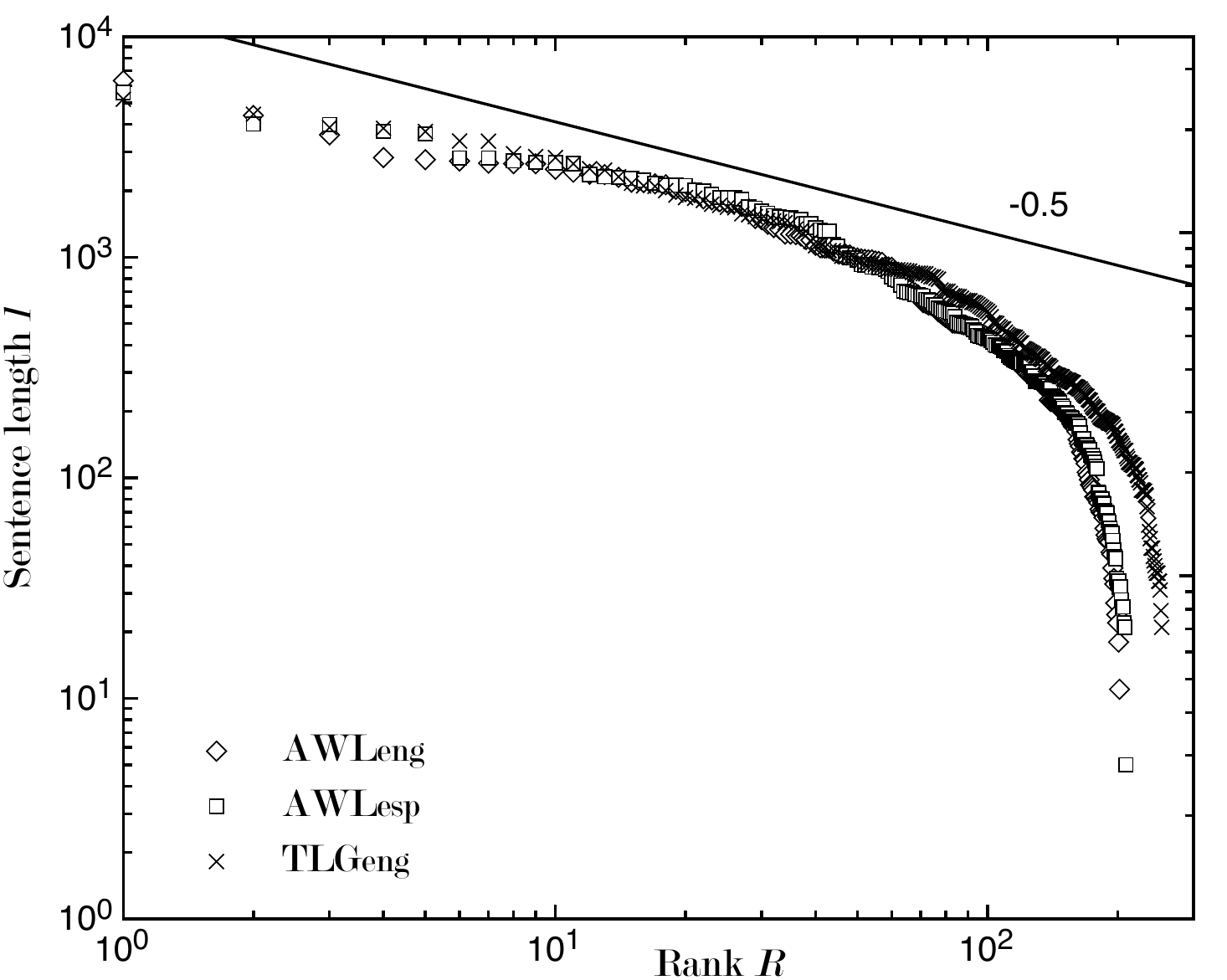}
\vspace{-0.01cm}
\caption{\label{5} Zipf (log-log) plot of the FTS of sentence lengths, as separated by (a) exclamation points, (b) question marks,  in the three texts of interest AWL$_{\mbox{eng}}$, AWL$_{\mbox{esp}}$ and TLG$_{\mbox{eng}}$. The 0.50 exponent of the corresponding Zipf law is indicated as a guide to the eye. A Zipf-Mandelbrot law fit for  various $R$ ranges  is not shown but is discussed in the main text; see also Table III}
\end{figure}

The (FTS)  Zipf original method  (though see \cite{yule})
\cite{Zipf1,15,16}  examines  the probability distribution of words in spoken (more exactly written) languages.
Zipf calculated the number   $N$ of occurrences of
each word in a given text. By sorting out the words according to their frequency
$f$, i.e. $N$ measured with respect to the total number $M$ of words in the text, a
rank $R$ can be assigned to each word, with $R=1$ for the most frequent one. For
any natural languages, one observes a power law  for the rank distribution
\begin{equation} f \sim R^{-\zeta} \label{eq-1} \end{equation}
with an exponent $\zeta$ close
to unity. The occurrence of this power law  has already been suggested \cite{zipfregimes} to be due to the ''hierarchical structure'' of the text as well as the presence of long range correlations (sentences, and logical structures therein).
 This strong quantitative statement with ubiquitous applicability is attested over a vast repertoire of human languages \cite{west}. Yet it is of empirical evidence that Zipf' s law in this (FTS) form   can at most
account for the statistical behaviour of words frequencies in a  zone spanning  the middle-low to low range of the rank variable. Even in the case of long single texts ZipfÕ s law renders an acceptable $\zeta$ in the small window between $s\simeq$ 10 and 1000, which does not represent a significant fraction of any literary vocabulary. However power laws lead to valuable  insights into  statistical processes, since they imply no scaling, whence  some hierarchical structure. The $\zeta$ exponent, or more generally the exponent of such a power law, can be turned into a {\it fractal dimension} (or Hurst exponent)  interpretation as in \cite{interpretationzD}.

 One difficulty stems in the lower and upper ranks of such plots because of the
 abundance and rarity of words \cite{rr1}.  Mandelbrot \cite{3,bbm2,bbm1} using arguments based on
   $fractal$ ideas, applied to the structure of lexical trees, improved the original form of the law, writing,
in terms of   two parameters $A$ and $C$ that need to be adjusted to the data,

\begin{equation}
f(R)= \frac {A }{(1 + CR)^{\zeta^*} }.
\label{eq-2}
\end{equation}

The latter form is thought to be more adequately valid for many sorts of data in the region corresponding to the $lowest$ ranks, that is $R <$ 100, dominated by mostly (small) function words.  In the same spirit one can show that  $w(h) \sim 1/h^{1+\nu}$ \cite{metois4};  whence $\gamma= 1+\nu$,  with $\nu= 1/\zeta$ or $ \zeta^*$ \cite{zanette,montemuro}. We do not discuss further the validity of Zip law(s) for which there is an abundant literature \cite{httpZipf}.

It has been shown that this  Zipf-Mandelbrot law is also obeyed by so many random processes  \cite{4,5} that one has been sometimes ruling out any interestingly special character for $linguistic$ studies. Nevertheless, it has been argued that it is possible to discriminate between human writings \cite{vilenski} and stochastic versions of texts 
precisely by looking at statistical properties of words that fall  where
Eq.(\ref{eq-1})  does not hold \cite{LTSpanos}.  Whence still some question cannot be avoided on artificial languages, translations, and on effects resulting from automatic or machine translations \cite{kanter}. 

Flipping the horizontal and vertical axes of the log-log plot suggested by Eq. (1) the cumulative probability distribution function (cPDF) $P(f)$ of the quantities of interest obeys

 \begin{equation}
P(\ge f) \sim f^{1-\eta} 
\label{eq-3}
\end{equation} 
where $1-\eta$ is a characteristic power law exponent for the cPDF. Whenceforth, $\eta-1$ = $1/\zeta$; i.e. $p(f) \simeq f^{-\eta}$. 




Note that in the following  the length of sentences is  also examined from the point of view of the numbers of characters    between  (six sorts of) punctuation marks, see Table 1.  One also distinguishes between the first and second chapter, thereby allowing for some consistency test.

\begin{table}
\tiny{
\begin{center}
\begin{tabular}{|c|c||c|c|} \hline
AWL$_{\mbox{eng}}$ & $f$ & AWL$_{\mbox{esp}}$ & $f$ \\ \hline \hline
the	&	1527	&	la (=the) &	2070	\\ \hline
and	&	802	&	kaj (=and)	&628	\\ \hline
to	&	725	&	\^s i (=she)	&508	\\ \hline
a	&	615	&	ne (=no/not)	&426	\\ \hline
I	&	545	&	mi (=I)	&403	\\ \hline
it	&	527	&	Alicio (=Alice)	&347	\\ \hline
she	&	509	&	diris (=said)	&332	\\ \hline
of	&	500	&	al (=to)	&313	\\ \hline
said	&	456	&	vi (=you)	&302	\\ \hline
Alice	&	395	&	ke (=that)	&292	\\ \hline
\end{tabular}
\end{center}
\caption{Top ten most frequent words in AWL$_{\mbox{eng}}$ and AWL$_{\mbox{esp}}$ with their frequency}
}
\label{tab-2}
\end{table}


 \subsection{Grassberger-Procacia method}

 \begin{figure}
\hspace{-0.1cm}
\includegraphics[width=2.5in]{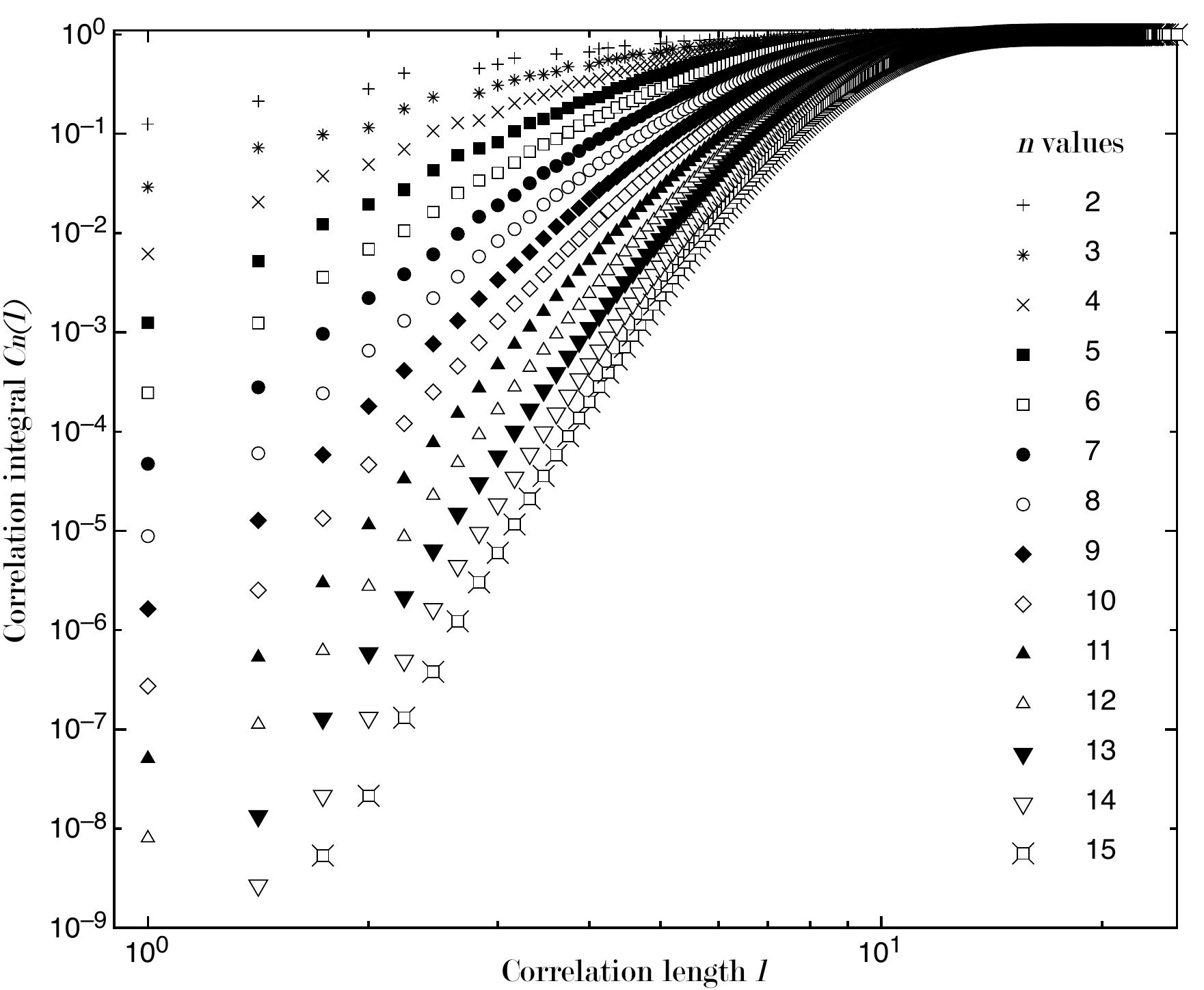}
\vspace{-0.01cm}
\includegraphics[width=2.5in]{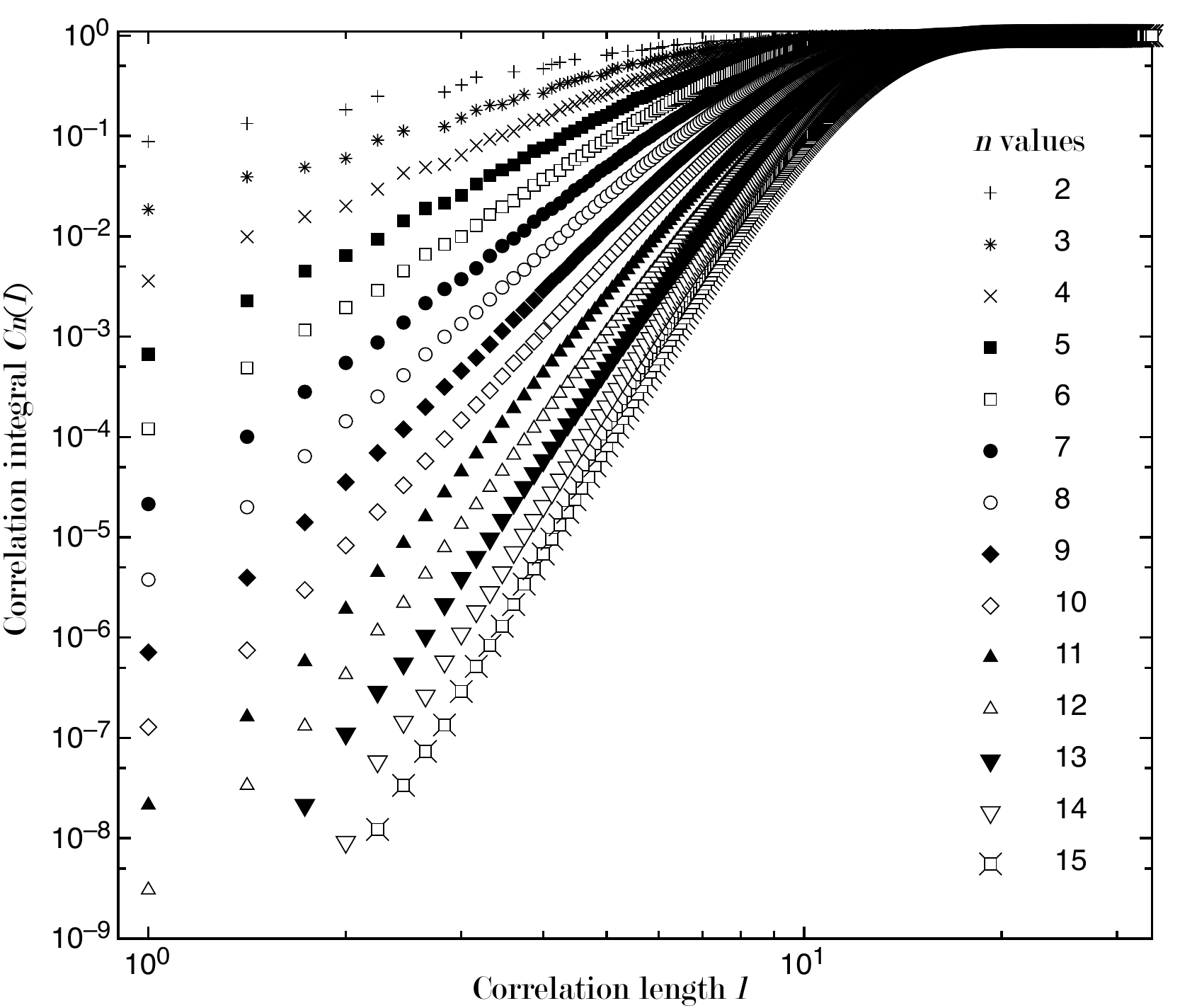}
\vspace{-0.01cm}

\caption{\label{6} Grassberger-Procaccia (log-log) plot of the correlation integral $C_{n}(l)$ as a function of the correlation length $l$  in different phase space dimension $n$, see text,     in the two texts of interest  (a) AWL$_{\mbox{eng}}$ and (b) AWL$_{\mbox{esp}}$ }
\end{figure}

      In order to get an insight into the dynamics of a system solely from the knowledge of the time series,  a
method derived by Grassberger and Proccacia \cite{GPprl,GPphd} has been proven to be particularly useful. This method has been applied to analyze the dynamics of 
neural network activity \cite{36}, electric activity of semiconducting circuits \cite{37,38},  climate \cite{GPclimate}  
, etc.

We  aim to finding some answer to  questions like 
\begin{enumerate} 
\item  Can the salient features of the system be viewed as the manifestation of a deterministic dynamics, or do they contain an irreducible stochastic element?
\item  Is it possible to identify an attractor in the system phase space from a given time series \cite{dimension}?
\item  If the attractor exists, what is its dimensionality $r$ \cite{theiler2}?
\item What is the (minimal) dimensionality  $n$ of the phase space within which the above attractor is embedded \cite{takens}? 
\end{enumerate}
This defines the minimum number of variables that must be considered in the description (through some model) of the  system. 

This is done as follows:
Let the LTS time series having $M$ data points, i.e. $y_i$ ($i = 1, \ldots , M$). Consider  the data as illustrating some dynamical process in a (phase) space with dimension $n$. Construct a set of  $V$ vectors $v_k$ ($k = 1, \ldots, V$) containing $n - 1$ points as follows:

\begin{equation}
v_k = (y_k, y_{k+\tau} , y_{k+2\tau} , \ldots, y_{k+(n-1)\tau}) 
\label{eq-4}
 \end{equation}
 where $\tau$ is an integer, called the {\it delay time}. Notice : $V +(n-1) = M$. In other words, one considers $k +n\tau$ as a sum   modulo $M$. Next one estimates the correlation integral from  the $distance$ $|v_i - v_j | $ between all the vectors such that $ 1 \leq i, j \leq V$. The correlation integral  $C_n(l)$ is obtained from 
 
 \begin{equation}
 C_n(l) =  \frac{ \mbox{\# of pairs $(i, j)$ such that  $|v_i - v_j | < l$}}{ N^2}.
 \label{eq-5} 
 \end{equation}

In  other words, 
 \begin{equation}
 C_n(l) =  \frac{\mbox{\# of pairs $(i, j > i)$ such that $|v_i - v_j | < l$} }{N(N-1)/2}.
\label{eq-6} 
 \end{equation}
 
 GP have shown that for small $l$, one has 
 \begin{equation}
  C_n(l) \simeq B l^r . \label{eq-7}
\end{equation}

where $B$ is some constant and $r$ is the so called {\it attractor  (correlation) dimension}, measuring the number of dynamic variables  or number of degrees of freedom. In order to obtain $r$ for  the different $n$ values, a log-log plot is in order. The choice of $\tau$ is debatable\cite{debatable}. In the following we have chosen $\tau= 500 $ like in other related studies \cite{LTSpanos}, for $n$ = 1 to 15. 

Practically, it was noticed that the correlation integral calculated for  $|v_i < v_j |$  distances takes a finite number of values; therefore each distance $l$ was  ''measured'' up to three decimal digits. Therefore two distances differing by less than 0.001 are not differentiated. Even though we have not tested the robustness of this  ''numerical approximation'',  we have not the impression that it is a drastic one.

A fit of the beginning of the $C_n(l)$ evolution through the best mean square technique on a log-log plot leads to a value of the relevant slopes, thus $r$ defined by 
Eq. (\ref{eq-7}).

 \section{Results}

\subsection{Zipf plots}

The result of the FTS analysis for the three main texts is shown in Fig. \ref{1}.   Figs. \ref{2}(a-b) show the (frequency, rank) relation for the esperanto and english chapter 1 and 2 respectively of both AWL texts. Each log-log plot roughly  indicates a linear relationship, for $R\ge10$, thus a $\zeta$ exponent close to unity, as often found, in usual literature. Some curvature is found for all texts below $R\sim15$ where a so called discontinuity exists, explained by \cite{powers} as due to  a transition between colloquial (''common'') small  and ''distinctive'' words. Some break, or change in slope,  is also found ca. 100, - see discussion in \cite{powers}. More interestingly let it be observed that the Rank =1 for the esperanto text is much higher than for the english texts. Moreover the  variety of distinct words is larger in esperanto as well. In between the number of words is less frequent in general, indicating a greater simplicity in vocabulary. The same is true whatever the chapter considered.

The top ten most frequent words in AWL$_{\mbox{eng}}$ and AWL$_{\mbox{esp}}$ are given with their frequency in Table \ref{tab-2}. It seems of interest to point out differences in style appearing from such a table. Notice that a translation does not conserve the number of words in a text, nor their importance in frequence. Of course the ranking might be intrinsically different, but also the translator can modify some sentences according to the language and grammar. An interesting illustration is noticed in Table \ref{tab-2}  :  the same words  \textit{'the'='la'} and \textit{'and'='kaj'} are both times the most frequent, but for example \textit{'Alice'} occurs more frequently in the english text than in the esperanto text, same for \textit{'I'='mi'}, even though \textit{'she'='\^si'} occurs an equal number of times. 

As indicated in the main text, one can  look at the length of sentences, in the case of the three main texts, Figs. \ref{3}-\ref{5}.  The relevant separators are mentioned in the figure captions. A marked difference occurs between the cases ``.'' (dot) and ``,'' (comma) on one hand and the others, column, semi-column, exclamation point, question mark, i.e.,  ``:'', ``;'', ``!'', ``?''.
In the first group, the slope is rather close to 1/3, but is closer to 1/2 for the latest four cases. The number of punctuation marks is relatively equivalent (see Table \ref{tab-1}) in all texts, but the number of dots and commas are much larger than the other punctuation marks. Therefore one might expect some finite size effect. Whence it is of interest to test Eq.(\ref{eq-2}), and compare the parameter values, given in Table \ref{tab-3}.

\begin{table}
\tiny{
\begin{center}
\begin{tabular}{|c|c||c|c|c|cr|} \hline
Fig. & Text & A & C & $\zeta^*$ & \multicolumn{2}{c|}{Range} \\ \hline
& & & & & \multicolumn{2}{c|}{$\ldots \le R\le \ldots$} \\ \hline
1	& AWL$_{\mbox{eng}}$ & 1177 & 0.17 & 1.15 & 2 ... &... 1000 \\
	& AWL$_{\mbox{esp}}$ & 962 & 0.28 & 1.01 &2 ... &...  1000 \\ 
	& TLG$_{\mbox{eng}}$ & 1098 & 0.13 & 1.21&2 .... &... 1000  \\ \hline \hline
2a & AWL$_{\mbox{eng}}$ & 116 & 0.19 & 1.16 &1... &... 200 \\
	& AWL$_{\mbox{esp}}$ & 48 & 0.15 & 0.01 & 4 ... &...  200 \\ \hline \hline
2b& AWL$_{\mbox{eng}}$ & 118 & 0.24 & 1.07 & 1 ... &...  200 \\
	& AWL$_{\mbox{esp}}$ & 168 & 0.90 & 0.92 & 2 ... &...  200  \\ \hline \hline
3a   & AWL$_{\mbox{eng}}$ & 1062 & 0.08 & 0.55 & 4 ... &...  200 \\
	& AWL$_{\mbox{esp}}$ & 984 & 0.5 & 0.36 & 1 ... &...  200 \\ 
	& TLG$_{\mbox{eng}}$ & 1029 & 0.46 & 0.34 & 1 ... &...  200  \\ \hline \hline
3b   & AWL$_{\mbox{eng}}$ & 366 & 0.09 & 0.33 & 1 ... &...  200 \\
	& AWL$_{\mbox{esp}}$ & 1019 & 1.2 & 0.34 & 4 ... &...  200 \\ 
	& TLG$_{\mbox{eng}}$ & 382 & 0.11 & 0.27 & 1 ... &...  200  \\ \hline \hline
4a   & AWL$_{\mbox{eng}}$ & 4650 & 0.06 & 1.15 & 2 ... &...   100 \\
	& AWL$_{\mbox{esp}}$ & 6978 & 0.14 & 0.97 & 1 ... &...  100 \\ 
	& TLG$_{\mbox{eng}}$ & 13128 & 0.02 & 4.73 & 1 ... &...   60  \\ \hline \hline
4b   & AWL$_{\mbox{eng}}$ & 5068 & 0.34 & 0.59 & 1 ... &...  100 \\
	& AWL$_{\mbox{esp}}$ & 5645 & 0.3 & 0.6 & 1 ... &...  100 \\ 
	& TLG$_{\mbox{eng}}$ & 3296 & 0.05 & 0.94 & 1 ... &...  100  \\ \hline \hline
5a   & AWL$_{\mbox{eng}}$ & 2756 & 0.03 & 1.48 & 4 ... &...  200 \\
	& AWL$_{\mbox{esp}}$ & 3630 & 0.02 & 1.91 & 3 ... &...  200 \\ 
	& TLG$_{\mbox{eng}}$ & 4777 & 0.39 & 0.61 & 1 ... &...   200  \\ \hline \hline
5b   & AWL$_{\mbox{eng}}$ & 3283 & 0.01 & 3.61 & 4 ... &...  100 \\
	& AWL$_{\mbox{esp}}$ & 4099 & 0.02 & 1.89 & 2 ... &...  100 \\ 
	& TLG$_{\mbox{eng}}$ & 5504 & 0.13 & 0.84 & 1 ... &...  100  \\ \hline
\end{tabular}
\end{center}
\caption{Values of parameters for the Z-M fit, Eq. (2); the corresponding figure and texts are  indicated together with the fit range}
}
\label{tab-3}
\end{table}

From the figures one can notice that three exponents seem to characterize the rank law 1.0, 1/3,  1/2.
The unity is $usual$. To find low values like 0.5 and 0.33 is more rare. Let us refer to a case in which it was striking to find a value ca. 0.55 and 0.72, i.e. the case of city size, in 1600 and 1990 respectively, and  0.74 for firm sizes with more than 10 employees in 1997, in Denmark \cite{knudsen}, in contrast to the usual 1.0 value \cite{gabaix}. A value smaller than unity indicates a more homogeneous repartition of the variables (words, here). One can see some analogy between city and firm sizes from the point of view of flow in (and out) of citizens or assets. Whence a Gabaix \cite{gabaix} or Simon \cite{simon} model can be  thought of to understand the values found here.  E.g. Gabaix claims that  two causes can lead to a value less than 1.0, i.e.  either :
\begin{enumerate}
 \item  the mean or variance of the growth process deviates from Gibrat's law \cite{gibrat}, i.e. the   growth rate  is independent of the size,
 or
 \item    the variance of the growth process depends is size-dependent.
\end{enumerate} 

Recalling that one does not examine the ''growth'' of the text at this stage yet, nor have any model for  doing so presently,  - except that of Simon \cite{simon} (words not yet used are added at a constant rate, while words already used are inserted at a frequency depending of the previous number of occurrences; this leads to Zipf law;  thus the rate of appearance of new words in fact decreases as the text length increases),
 one can nevertheless agree that the sample size is relevant for finding a small $\zeta$ value. Indeed it is clear that the found values correspond to the length of sentences which are defined through various punctuation marks, counting characters rather than words. Several orders  of magnitude in the maximum rank distinguish the cases. What is still surprising is why the  longest sentences, thus defined through dots  and commas lead to  a smaller values than for other punctuation marks which lead to less frequent sentences. 

An alternate view can be taken through the Z-M analytical form, Eq.(2).  Values of the parameters are given in Table 3 for various ranges $R$. it is fair to state that the parameters are NOT very robust with respect to the range. Values of $\zeta^*$ can be found close to the apparently best looking slope, $\zeta$, but other values can be found as well. This is due to the strong influence of the low rank points. The paradoxical situation occurs when one remembers that the analytical form is supposed to be used in order to take into account the finite value at $R= 1$. However the curvature for the (small) function words markedly influences the outcome. In order to illustrate the point, a brief example is given in an Appendix.

\subsection{Grasseberger-Procaccia plots}

 The analysis of correlation integrals allow to estimate whether the number of degrees of freedom (of a process) is large or reasonably small. It seems that the usual goal is rather qualitative. However it pertains to the fundamental question on noisy signals, - is it noise or chaos?  As explained here above the algorithm is based on the statistics of pairwise distances for an arbitrary choice of the delay time. Therefore the output of the method  results in observing an evolution of correlations, i.e. in the knowledge on how often a point in some ((= ''the'') phase space is found near another, whence illustrating some dynamical features connecting local and global features.
 
  The three sets of correlation integrals, calculated following the method here above described,  are shown in Figs. 7-8.  The slopes  can be summarized through a graph relating $r$ and $n$ (Fig.9). It is found that  the attractor dimension $n$ is not only smaller than the space dimension, as it should be \cite{takens}, but  also is  a linear  function on a log-log plot  of the so called phase space dimension $r$, {\it for the three texts of interest}. A remarkable power law  is found, whatever the text, $r = n^{\lambda}$, with $\lambda = 0.79 $, which does not indicate any saturation. It seems of great interest to examine other authors and to find whether $\lambda  $ characterizes some style or author or ....



\begin{table}\tiny{
\begin{center}
\begin{tabular}{|c||c|c|} \hline
Text & Slope $r(n)$ & Standard deviation  \\ \hline \hline
AWL$_{\mbox{eng}}$	&	0.84	&	0.01	\\ \hline
AWL$_{\mbox{esp}}$	&	0.747	&	0.004	\\ \hline
TLG$_{\mbox{eng}}$	&	0.77	&	0.01 \\ \hline \hline
Global &	0.79 & 0.01	\\ \hline
\end{tabular}
\caption{Measured slopes of the linear function $r(n)$}
\label{tab-4}
\end{center}
}
\end{table}

 \begin{figure}
  \begin{center}
\hspace{-0.5cm}
\includegraphics[width=3.5in]{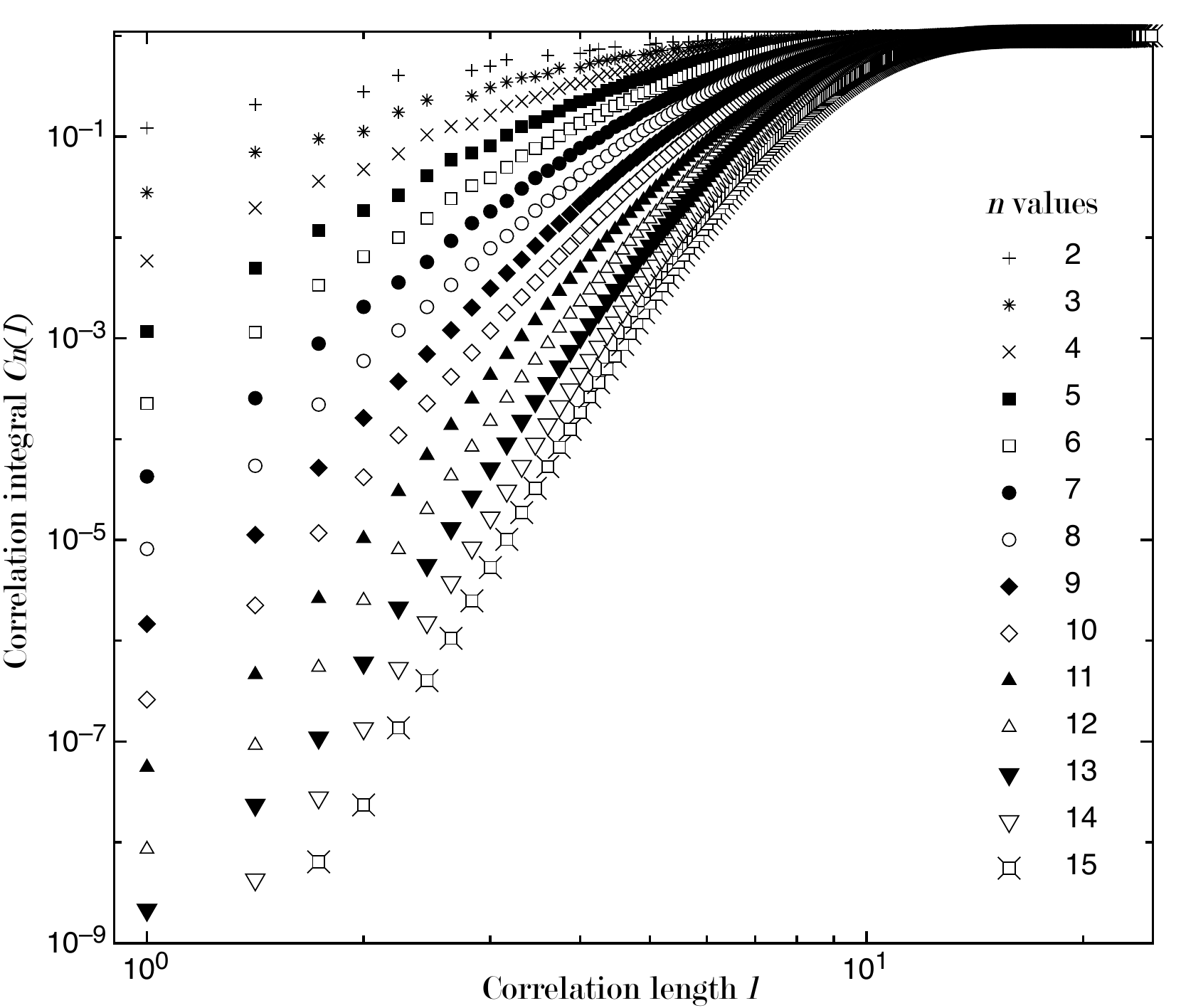}
\caption{\label{7}  Grassberger-Procaccia (log-log) plot of the correlation integral $C_{n}(l)$ as a function of the correlation length $l$  in  phase spaces  with different dimensions ($n$)     for   TLG$_{\mbox{eng}}$ }
\end{center}
\end{figure}

 \begin{figure}
  \begin{center}
\hspace{-0.5cm}
\includegraphics[width=3.5in]{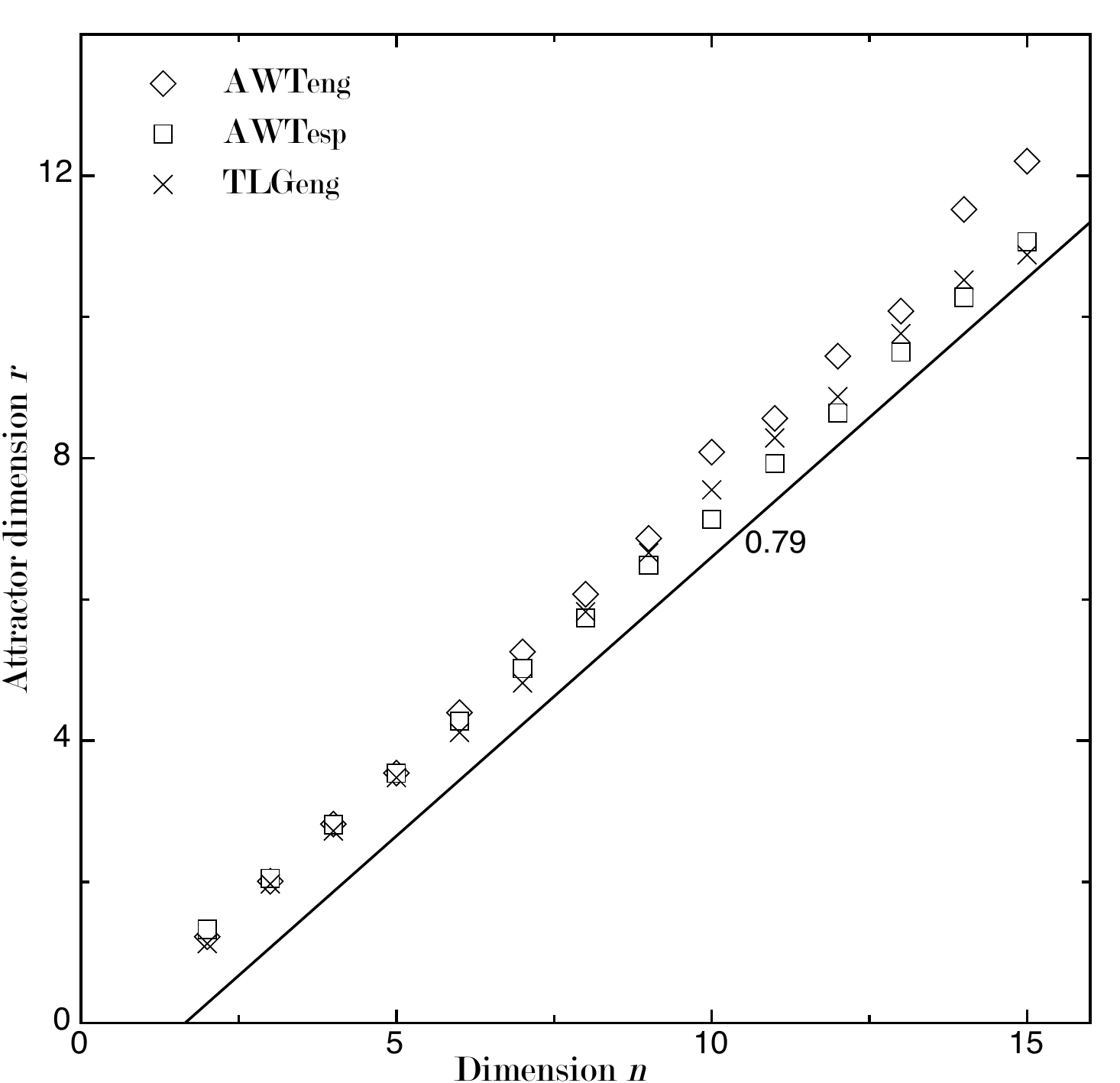}
\caption{\label{8} Attractor dimension $n$ as a function of the so called phase space dimension $r$ for the three texts of interest. Notice that a linear relationship is found with a proportionality coefficient $ \lambda $= 0.79 on a log-log plot, as for $r = n^{\lambda}$}
 \end{center}
\end{figure}

\section{Conclusion}

   At first sight, a time series of a single variable appears to provide a limited amount of information. We
usually think that such a series is restricted to a one-dimensional view of a system, which, in reality, contains a
large number of independent variables. On one hand  FTS and LTS  result from a dynamical process, which is usually first characterized by its fractal dimension. The first approach should contain a mere statistical analysis of the output, as done through a Zipf like analysis. It can be found that analytical forms, like  power laws   with different characteristic exponents for the ranking properties exist.  The Zipf exponent can take values $ca.$ 1.0, 0.50 or 0.30, depending on how a sentence is defined.  This non-universality is conjectured to be a measure of the author $style$. Another approach through a Zipf-Mandelbrot law seems unreliable due to the (present lack) of distinction most likely between function and determining words, and breaks occurring in the $f(R)$ plots. Something which has not been examined and is left for further studies is the  distinction between oral-like and descriptive parts of a tgext and its translation.

Moreover  a time series is known   \cite{GPprl,GPphd,GPclimate}  to bear the marks
of all other variables participating in the dynamics of the system. Thus one is  able to ÔÔreconstructÕÕ the
systems phase space from such a series of one-dimensional observations.
 When applying the Grassberger-Proccacia (GP) method to a physics time series one wants to know whether the attractor is based on a finite set of variables. The lack of saturation  found here through the law
 $r = n^{\lambda}$
for the size of the attractor indicates that the writing of a text by some creative author can be hardly reduced to a finite set of differential equations !   Yet the analytical form suggests to examine whether $\lambda$ characterizes an author style or creativity, and how robust its value can be.

Finally, as in \cite{LTSpanos} we concur that the application of GP analysis indicates that linguistic
signals may be considered as the manifestation of a complex system of high dimensionality, different from random signals
or systems of low dimensionality such as the Earth climate or financial signals.

Last but not least as on comparing AWL$_{\mbox{eng}}$,  AWL$_{\mbox{esp}}$ , and TLG$_{\mbox{eng}}$, with both the ''static'' and ''dynamic'' methods, it seems that the texts are qualitatively similar, which indicates ... the quality of the translator. In this spirit, it would be interesting to compare with results originating from text obtained through  a machine translation, as recently  studied in  \cite{koutsoudas}. It is of huge interest to see whether a machine is {\it more flexible} with vocabulary and grammar than a human translator, - see also \cite{kanter}!
 


\bigskip{}

 {\it \bf Acknowledgements}

The author  would like to thank D. Stauffer for usually fruitful discussions.
This work  has been partially supported by European Commission Project 
CREEN FP6-2003-NEST-Path-012864 and
European Commission Project 
E2C2 FP6-2003-NEST-Path-012975   (E2C2: Extreme Events: Causes and Consequences).

This paper results from the work of Mr.  Jeremie  Gillet, when an undergraduate student attending my lectures on fractals in 2006-2007. However his present Ph.D.  advisor forbids him to coauthor a paper with me. I am very  thankful to JG to allow me to publish the results of his data analysis and for many comments. 

\section{Appendix}

\begin{table}
\tiny{
\begin{center}
\begin{tabular}{|c||c|c|c|} \hline
AWL$_{\mbox{eng}}$  &parameter & value & absolute error\\ \hline \hline
Range :  from 2 to 200
&A        &        2104.5665             &        102.1277\\ 
&C&	 1.2151	&	0.2201	\\ 
&$\zeta^*$	&	 0.3924    	&	6.1668e-3	\\ \hline
Range :  from 2 to 200
&A        &        1239.7700              &        18.5571\\ 
&C         &           0.1553                   &         1.1255e-2\\ 
&$\zeta^*$     &           0.4874               &         7.7509e-3\\ \hline
Range :  from 3 to 200
&A        &        1105.3531                  &        10.7540\\ 
&C         &           9.4509e-2             &         4.7342e-3\\ 
&$\zeta^*$        &           0.5334                  &         6.9828e-3\\ \hline
Range :  from 4 to 200
&A        &        1061.8008                   &        9.7680\\ 
&C         &           7.9410e-2             &        3.7663e-3\\ 
&$\zeta^*$     &           0.5526                 &        7.1248e-3\\ \hline
\end{tabular}
\end{center}
\caption{Effect of low ranking points on Z-M fit; parameter values and their corresponding error bar for  AWL$_{\mbox{eng}}$ sentences limited by ''dots''}
\label{tab-4A}
}
\end{table}

The Zipf-Mandebrot, Eq.(2), 	law is thought to be useful for better describing the ranking function $f(R)$, in particular in order to take into account the finite value of $f$ at $R \simeq 1$. Yet from data presented in Table 3, it can be observed that the parameters, in particular $\zeta^*$ is far from robust when the range of $R$ slightly varies. For example $\zeta^*$ can vary from 0.84  to  3.61 when only the fit range is slightly changed, like in the case of Fig. 5b,  for sentences limited by question marks in the three original texts where one expects an exponent near 0.5. It appears that if one fits from $R$=
  1 one finds $\zeta^*$=0.65, but from $R$= 2 , $\zeta^*$=1.68, from $R$=3,  
$\zeta^*$= 2.68,  and from $R$=4 $\zeta^*$=3.61, as shown in Table 3. This is ''obviously'' due to the curvature of the data at low $R$.
Some other example pointing to the probable origin of the fit parameter value instability.
in AWL$_{\mbox{eng}}$ defined through dots in Fig.1 is given in Table 4, where a few results of small changes in ranges, removing one, two, three or four first points,  and the corresponding parameter  fits are given  with the (absolute) error bars.  

 I have not found much discussion of the matter in the literature, maybe because either the case is not frequent, or not examined. See nevertheless \cite{kanterkessler}  where it is suggested that $\zeta^*$ be interval dependent and increasing logarithmically with $R$. In the present case, it appears that   one can consider  the origin of the instability to reside in the''large'' variations of $f(R)$ at small $R$. In fact the curvature of $f(R)$ changes from convex to concave at small $R$. This leads to an instability in the set of least mean square fits.  This, in other words, is due to the number of regimes, changes in curvature, in the data. Powers \cite{powers}  (and later others like \cite{17}) had already noticed   that one should distinguish between small (function) words and large (determining) words, and pointed to the break, or change in slope at finite $R$ ($\sim $100).  A recommendation is in order : a visual scan of the data should be made before attempting a fit with Eq.(2), in order to observe the number of regimes, or the number of crossover points,  which might appear in the data. It is also of course  useless to attempt a fit with many more parameters, - one would need at least three per regime! Yet the understanding of the position of the crossover points might be of interest. Recall the remarkable papers on the position of the cross over points in detrended fluctuation analysis studies \cite{hu,chen}, related to a periodic background or trend  in time series. Such considerations would illuminate in the present context,  the language  quality level or an author style and creativity through a text ''background'' content..

\end{document}